\documentclass[iop]{emulateapj}		
\usepackage{epsfig}			
\usepackage{graphicx,color}		
\usepackage{amssymb}			
\usepackage{color}			
\usepackage{url}			
\usepackage{amsmath}			
\usepackage{rotating}			
\usepackage{float}			
\usepackage{textcomp}			
\usepackage{psfig}
\usepackage{dcolumn}
\usepackage{times}
\usepackage{tabularx}
\usepackage[english]{babel}
\usepackage{ragged2e}

\shorttitle{Association Of Plages With Sunspots}
\shortauthors{S. Mandal et al.}

\begin{document}

\title{Association of Plages With Sunspots: A multi wavelength Study Using Kodaikanal C\MakeLowercase{a}  $\scriptsize{{\textrm{II}}}$ K and Greenwich sunspot area  Data}

\author{Sudip Mandal$^{1}$,
Subhamoy Chatterjee$^{1}$,
Dipankar Banerjee$^{1,2}$}
 
\affil{$^{1}$Indian Institute of Astrophysics, Koramangala, Bangalore 560034, India. e-mail: {\color{blue}{sudip@iiap.res.in}}\\
$^{2}$ Center of Excellence in Space Sciences India, IISER Kolkata, Mohanpur 741246, West Bengal, India  \\}
  \justify
  
\begin{abstract}
   {Plages are the magnetically active chromospheric structures prominently visible in Ca $\scriptsize{{\textrm{II}}}$ K line (3933.67 \r{A}). A plage may or may not be associated with a sunspot which is a magnetic structure visible in the solar photosphere. In this study we explore this aspect of association of plages with sunspots using the newly digitized Kodaikanal Ca $\scriptsize{{\textrm{II}}}$ K plage data and the Greenwich sunspot area data. Instead of using the plage index or fractional plage area and their comparison with the sunspot number, we use, to our knowledge for the first time, the individual plage areas and compared it with the sunspot area time series. Our analysis shows that these two structures formed at two different layers are highly correlated with each other on a time scale comparable to the solar cycle. The area and the latitudinal distributions of plages are also similar to that of the sunspots. Different area thresholdings on the `Butterfly diagram' reveal that plages with area $\geq$4 arcmin$^2$ are mostly associated with a sunspot in the photosphere. Apart from this, we found that the cyclic properties change when different sized plages are considered separately. These results may help us to better understand the generation and the evolution of the magnetic structures in different layers of the solar atmosphere.}
\end{abstract}
\keywords{Sun: Sunspots--- Sun: Activity --- Sun: Oscillation}

\section{Introduction}

Sun is a magnetically active star with a dynamic atmosphere which varies on a time scale from seconds to years. Different solar features are basically the manifestations of the solar magnetic concentraions in different layers of the Sun. Plages are the chromospheric features which appear as bright patches on the solar disc when seen through Ca $\scriptsize{{\textrm{II}}}$ K line (3933.67~\r{A}) images, whereas sunspots are the dark photospheric features prominently visible in white light images. Sunspots and plages both vary periodically in a 11 year time scale known as `solar cycle'. Apart from that plages are found to be highly correlated with the location of the magnetic field concentrations \citep{2011ApJ...730...51S,2016ApJ...827...87C}, very similar to sunspots.

Satellite measurements over the past few decade revealed that the changes in solar irradiance happen over various time scales and it has a strong dependence on the various solar features present on the solar disc \citep{2004A&ARv..12..273F,2004ApJ...611L..57F}.
 It has been found that the total solar irradiance  is highly correlated with the fractional plage area or the plage index \citep{2010SoPh..264...31B}. Recently, \citet{2016SoPh..tmp..121B} found a strong correlation between the sunspot number and the Ca $\scriptsize{{\textrm{II}}}$ K emission index. Thus the study of long term plage data is not only of importance in connection to the solar irradiation variation study but also with the evolution of the solar magnetic fields and its cyclic changes.\\

Different observatories around the globe has been taking routine observations of the plages in Ca $\scriptsize{{\textrm{II}}}$ K line. Mount Wilson data series is one of such plage index time series in the world \citep{2010SoPh..264...31B}. Kodaikanal observatory in India has obtained daily photoheliograms of the Sun since 1904 to till 2007. This century long data has been recently digitized  \citep{2014SoPh..289..137P}. \citet{2016ApJ...827...87C} used this digitized data to identify the plages using an automated algorithm and generated a plage area time series.\\

In this paper, we use this data to find an association of the plages with the sunspots. We also use this data to find different distributions in the plage sizes and their latitudinal locations.\\

\section{Data Description }
In this study we have used the plage area time series obtained from the newly digitized Kodaikanal Ca $\scriptsize{{\textrm{II}}}$ K data. The complete time series covers more than 100 years of data (1904 -- 2007). Due to issues with the conditions of the photographic plates and also large number of missing days in the later half of the data [see Fig.~1b and Fig.~4 of \citet{2016ApJ...827...87C}], we chose to limit our analysis between the period of 1907 to 1965 which covers from cycle 14 (descending phase only) to cycle 19. For every detected plage form the daily Ca $\scriptsize{{\textrm{II}}}$ K images, we have the heliographic latitude, longitude (in degrees) and the area (in arcmin$^2$) information.\\
 
 For the sunspot area data we have used the Greenwich daily sunspot record, for the same duration, available in the website \url{https://solarscience.msfc.nasa.gov/greenwch.shtml}.

\section{Results}
\subsection{Yearly Averaged data and The Hemispheric Asymmetry}

\begin{figure}[!htbp]
\centering
\includegraphics[width=0.49\textwidth]{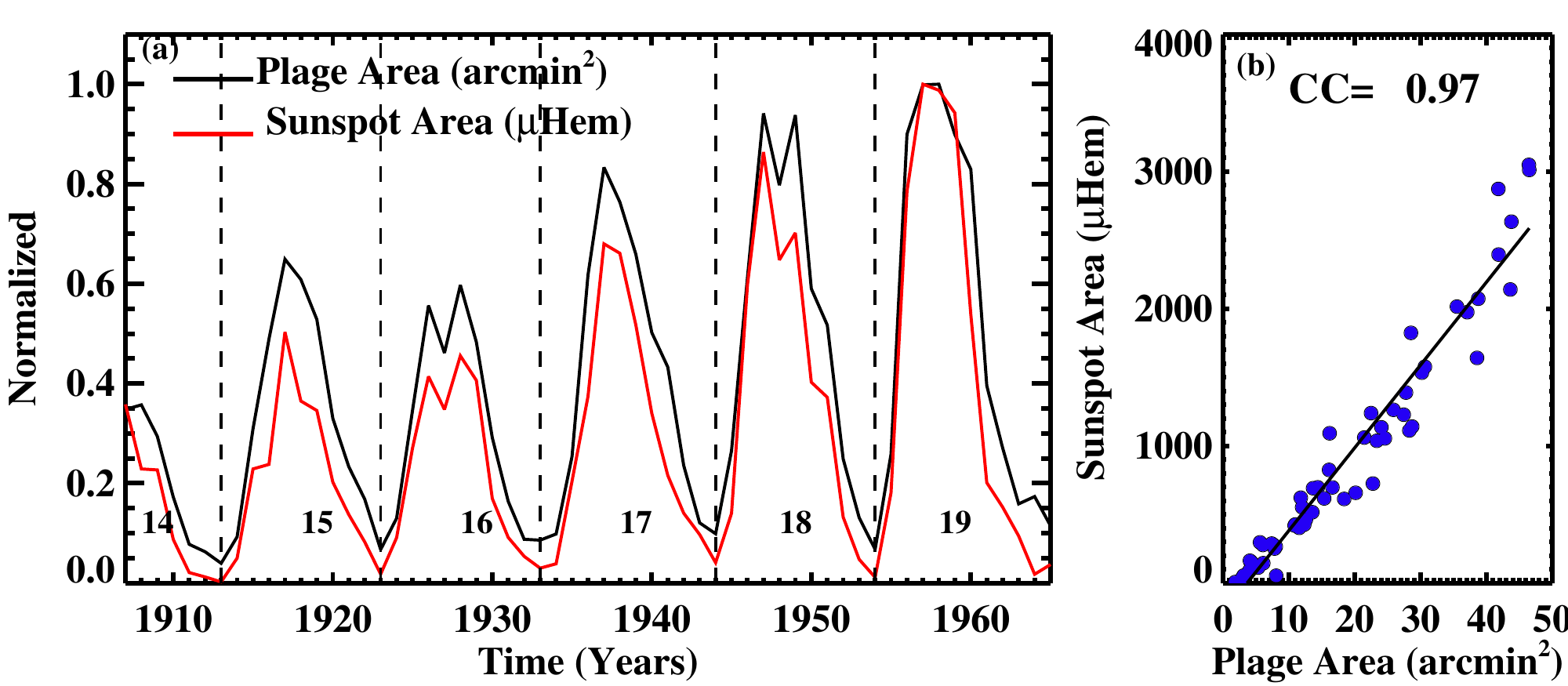}
\includegraphics[width=0.49\textwidth]{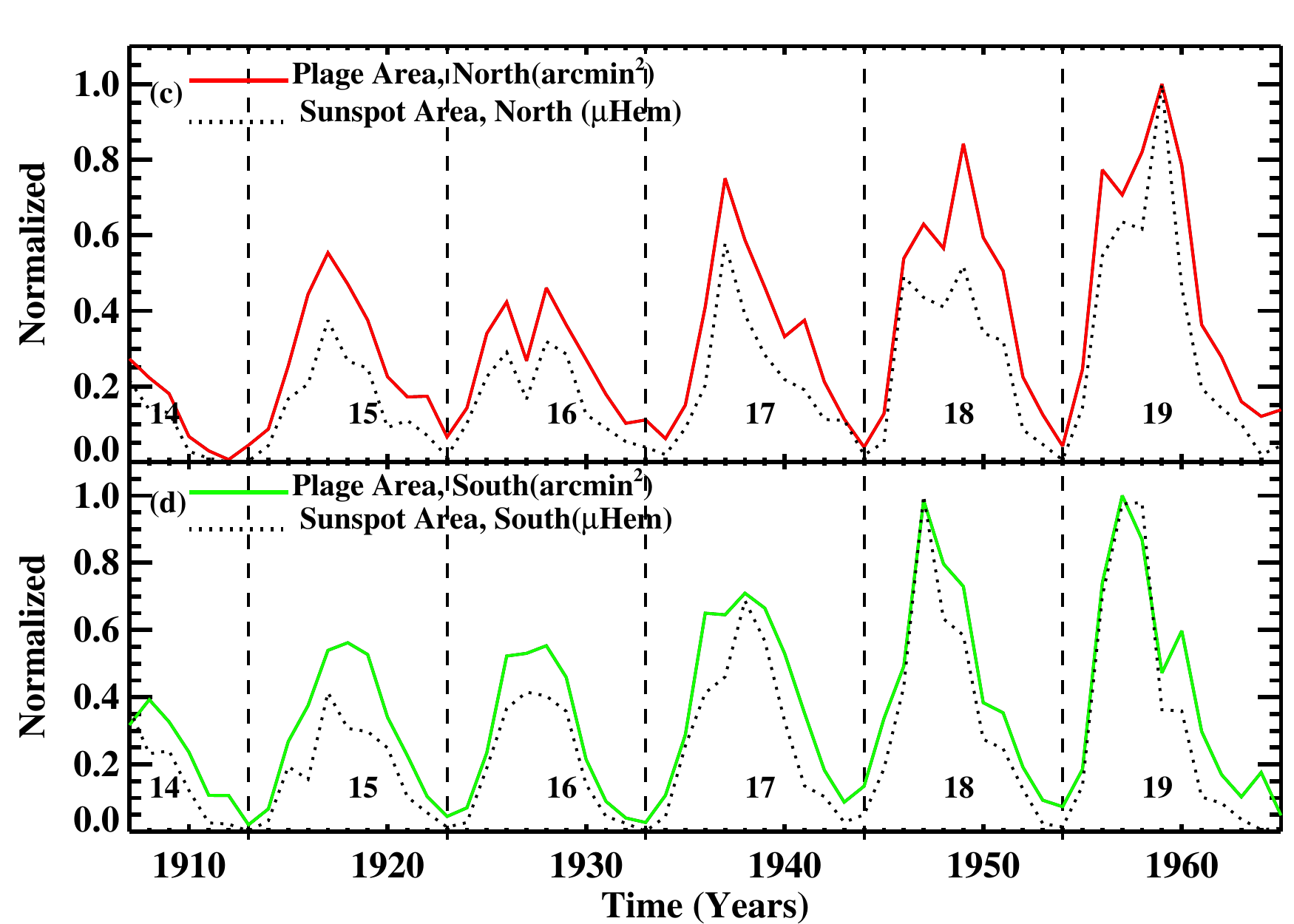}
\caption{ Panel (a) shows the comparison plot of the yearly averaged sunspot and the plage area data. Scatter plot between the two area data is shown in panel (b). Panels (c - d) show the   North (South)  hemispheric yearly averaged plage area as marked.
}
\label{yearly_comp} 
\end{figure}

We generate the yearly averaged data from the daily plage observations and plotted it (black solid line) in panel (a) of Figure~\ref{yearly_comp} along with the sunspot area data (red solid line). 
Since we are interested in the association of the two structures, we plot the normalized values of the yearly averaged data. From the figure we readily see that the two time series show a good match with each other. Every feature, including the double peaks seen in the sunspot data are also present in the plage area time series. To estimate this association quantitatively we plot the scatter diagram as shown in panel (b) of Figure~\ref{yearly_comp}. A correlation coefficient of 0.97 again confirms the close association of these two solar features which have formed in two different layers in the solar atmosphere. Though we must emphasize that this high correlation in the yearly data does not imply the same for the smaller time scales (in months or days). This is because when a sunspot decays away, the remnant small magnetic field may still continue to show itself as a plage in the higher atmosphere. Also the small scale magnetic fields, which lives in the order of days or less, do not always develop as a sunspot.\\

Now the hemispheric asymmetry in the sunspot area is a well known phenomena. We investigate the same from the plage area time series by computing the yearly data separately for the two hemispheres. Panels (c-d) in Figure~\ref{yearly_comp} show the yearly averaged plage area data in the northern and southern hemispheres (plotted in red and green solid line) respectively. The sunspot area data, for the corresponding hemispheres, is also overplotted in the panels as shown in the black dotted lines. Hemispheric plage area series shows a very good match with the sunspot area data. Similar to the sunspots, in this case too we find that the double peaks near the cycle maximum is not a persistent feature in both of the hemispheres. For an example, cycle 16$^{th}$ is a double peaked cycle (see Figure~\ref{yearly_comp}a) but from Figure~\ref{yearly_comp}(c-d) we see that in this case only northern hemisphere shows a double peak signature. In the case of 19$^{th}$ cycle, there is no such double peak seen in the overall case but both the hemispheres have prominent double peak signatures.\\
\begin{figure}[!htbp]
\centering
\includegraphics[width=0.49\textwidth]{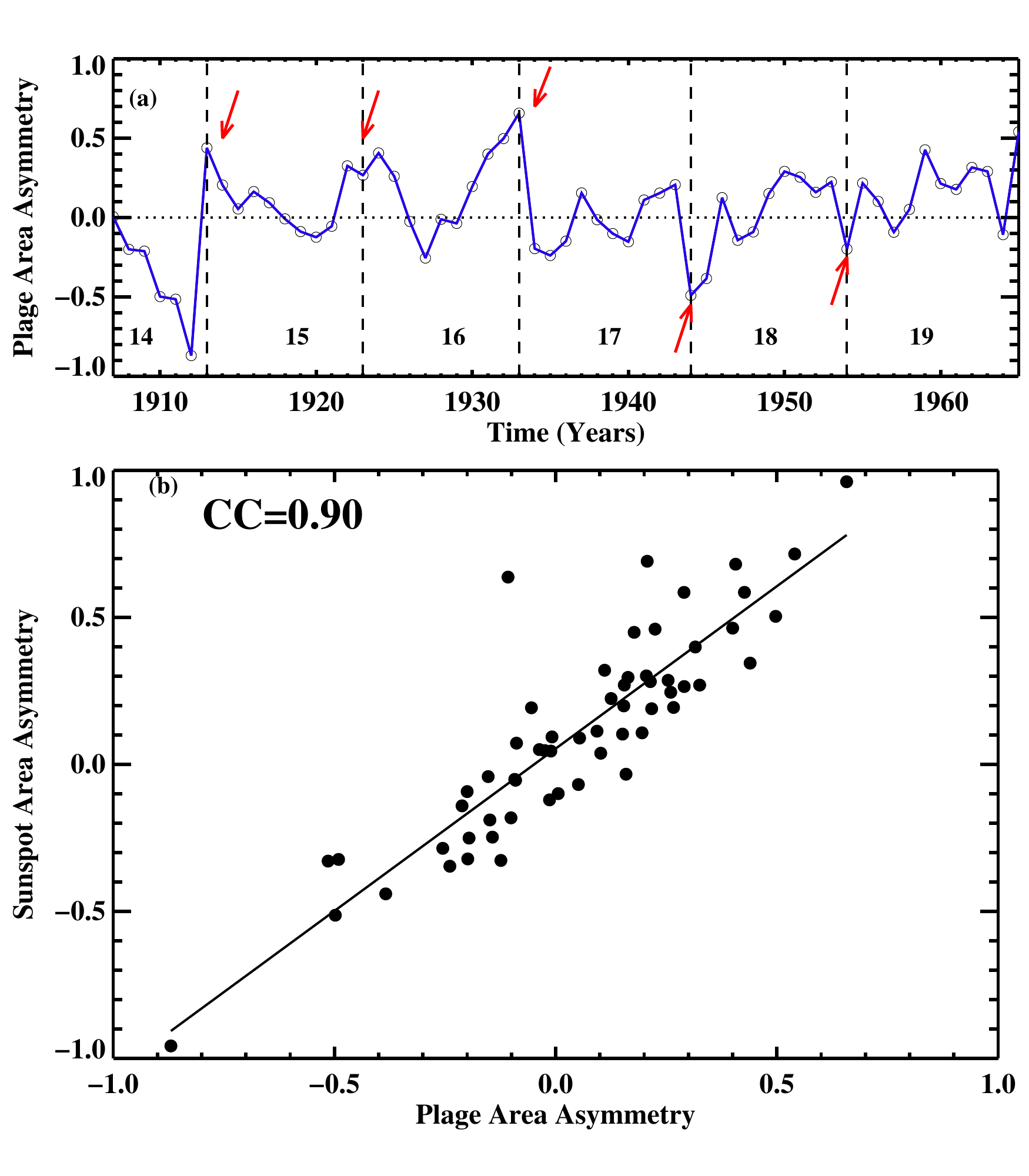}
\caption{ Normalized plage area asymmetry is plotted in panel (a). Comparison of the same with the sunspot area asymmetry is shown in panel (b). Correlation value between these two are printed in the panel.}
\label{plage_corr} 
\end{figure}

 Next we compute the normalized asymmetry coefficient, defined as (A$_{pn}$-A$_{ps}$)/(A$_{pn}$+A$_{ps}$), as a measure of the hemispheric asymmetry in the plage area (A$_{pn}$,A$_{ps}$ are the yearly averaged plage area values in the northern and the southern hemispheres). The evolution of this asymmetry coefficient is plotted in panel (a) of Figure~\ref{plage_corr}. There are quite a few distinct features readily noticeable from the plot. During the minima of cycles 14, 15 and 16, we see that the northern hemisphere dominates whereas for the later cycles, i.e cycles 17, 18 and 19, the southern hemisphere dominates over the north. This behavior is highlighted in the plot (Figure~\ref{plage_corr}a) by using red arrows. It is also interesting to note that during the progression of a cycle the asymmetry coefficient changes its sign quite a few times and this does not show any meaningful correlation with the cycle amplitude or any other properties of that particular cycle. We also notice that on an average the northern hemisphere dominates over the south for the six cycles analyzed here. We again revisited this northern hemispheric dominance in the following section.\\

Since we are interested in the plage-sunspot association, we look for the same in the asymmetry coefficient also. In panel (b) in Figure~\ref{plage_corr} we plot the scatter diagram between the asymmetry coefficients obtained from hemispheric plage area and the same computed for sunspot area. We find a very good match between these two coefficients with a correlation value of 0.90. Though we found some outliers also which probably have occurred during the cycle minima during which the asymmetry coefficient is prone to large departures.
\subsection{Plage Distributions}
\subsubsection{Size Distribution}
\begin{figure}[!htbp]
\centering
\includegraphics[width=0.49\textwidth]{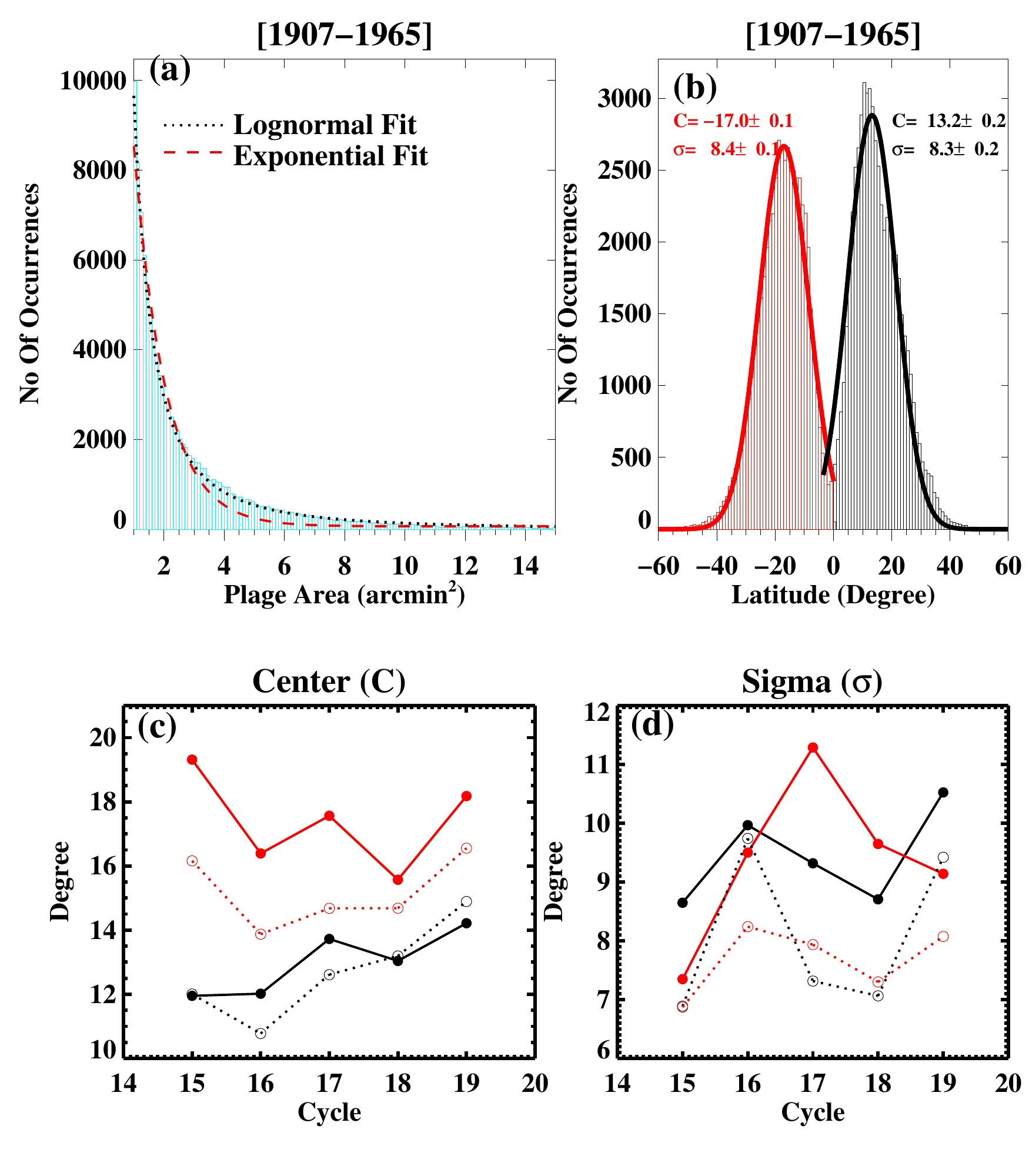}
\caption{ Panels (a) and (b) show the area and the latitudinal distributions of the plages. Corresponding fits to these distributions are also overplotted in these panels. Panels (c) and (d)  show (with solid black and red circles for the northern and southern hemisphere) the evolution of the fitted Gaussian parameters (C and $\sigma$) with the cycle number. We also overplot the same for the sunspots in open circles.}
\label{area_dist} 
\end{figure}

Individual plage area values are considered for the size distribution analysis. In panel (a) of Figure~\ref{area_dist} we show the plage area (sizes) distribution for the whole time period i.e from 1907 to 1965. The distribution pattern look similar to an exponentially decaying function. We fit the histogram with a decaying exponential function of the form, Y=A$_0\exp^{-(\frac{X}{B})}$, as shown by the red dashed line in Figure~\ref{area_dist}a. From the fit we notice that though the initial part of the histogram gets a good fit but the fitted function drops very rapidly in the wing and leaves a large deviation around that region.\\

To get a better fit, we consider the lognormal function next. This is inspired from that fact the sunspots are known to have a lognormal size distribution \citep{1988ApJ...327..451B,2005A&A...443.1061B, 2016arXiv160804665M}. Thus a lognormal function of the form, 
\begin{equation*}
\mathrm{y}=\frac{1}{\sqrt{2\pi}\sigma x}\exp{-\frac{[\ln(x)-\mu]^2}{2\sigma^2}} 
\end{equation*}
is considered and fitted the histogram as shown with the black dotted curve in Figure~\ref{area_dist}a. In this case we notice that the full histogram along with the tail gets a very good fit. Thus the individual size distribution of the plages also follow the same `lognormal distribution' as we find for the sunspots. Here we must highlight the fact that the good match of the two functions (exponential and the lognormal) at the core of the histogram (near to the origin, 1 arcmin$^{2}$) is due to the use of a rigid cutoff of 1 arcmin$^{2}$ as the minimum detectable plage area. Thus the initial increment of the lognormal distribution got suppressed and the function decays exponentially thereafter.\\ 
\subsubsection{Latitudinal Distribution}
  In the beginning of a cycle, plages occur at higher latitudes and progressively move towards the equator as the cycle progresses. We plot the distribution of the number of plages with their latitudes, for the full time span (1907-1695), in panel (b) of Figure~\ref{area_dist} (we also analyzed same for the individual cycles also). It shows two bell-shaped distributions corresponding to the two hemispheres which are then fitted with two separate Gaussian functions. From the plot (panel b) we notice that the peak height of the northern hemisphere is greater than that of southern hemisphere. This compliments our findings in the section 3.1, where we obtained a similar result from the area asymmetry analysis. To make a comparison, we repeat the same procedure for the Greenwich sunspots area data. From every Gaussian fit we note down two parameters: center (C) and the sigma ($\sigma$) values. In panels (c-d) in Figure~\ref{area_dist}, we plot the evolution of these parameters for the plages and the sunspots simultaneously (solid red and black circles represent the southern and northern hemispheric values for the plages whereas the open circles corresponds to sunspots).\\

In the case of the center (C) plot (panel c) we see that the the centers of the plage distributions, for the two hemispheres, is always higher than that of the sunspots, although the trends remain the same. At the same time we notice that the differences is more for the southern hemisphere. We also observe that the center of the southern hemisphere for the 16$^{th}$ cycle is higher than that of the 19$^{th}$ cycle though the cycle amplitude of the 19$^{th}$ cycle is much higher than that of the 16$^{th}$ cycle. In panel (d) we plot the evolution of the sigma parameters. Overall the evolution of this parameter for both the indices i.e for sunspots and the plages follow the same pattern. Again we find that there is a noticeable difference in the southern hemisphere. The maximum sigma value occurs for the 17$^{th}$ cycle whereas for the sunspot it occurs at 16$^{th}$ cycle. Currently there is a very little theoretical understanding on the relation of these parameters in connection to the solar dynamo. Recently \citet{2016A&A...591A..46C} found a connection of the parameter $\sigma$ with the diffusivity ($\eta$) parameter used in the dynamo theory.

\subsection{Correspondence Between Plage and Sunspot Locations}

\begin{figure*}
\centering
\hbox{
      \includegraphics[width=11.5cm,height=15cm]{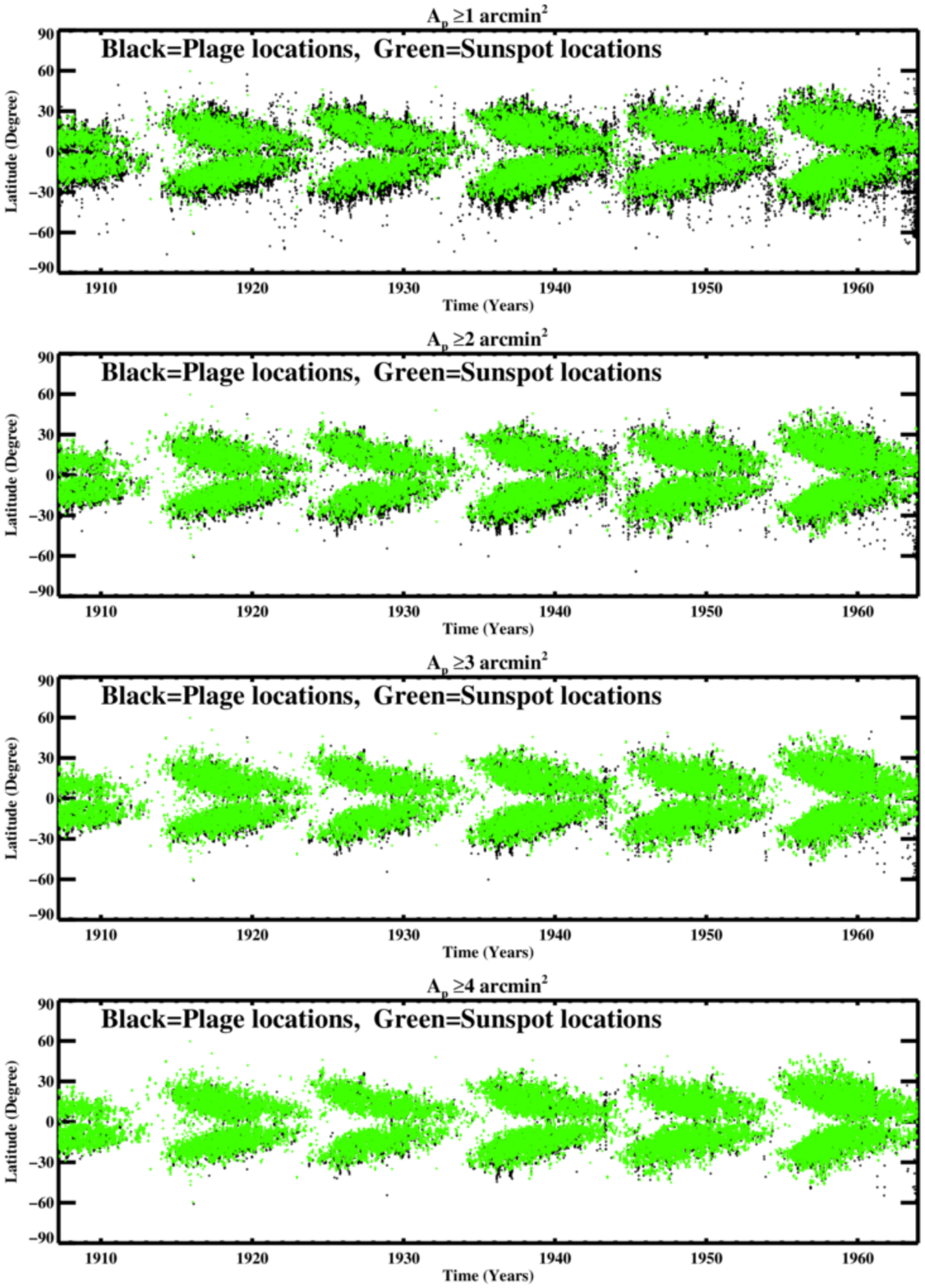}
      \includegraphics[width=4cm,height=15cm]{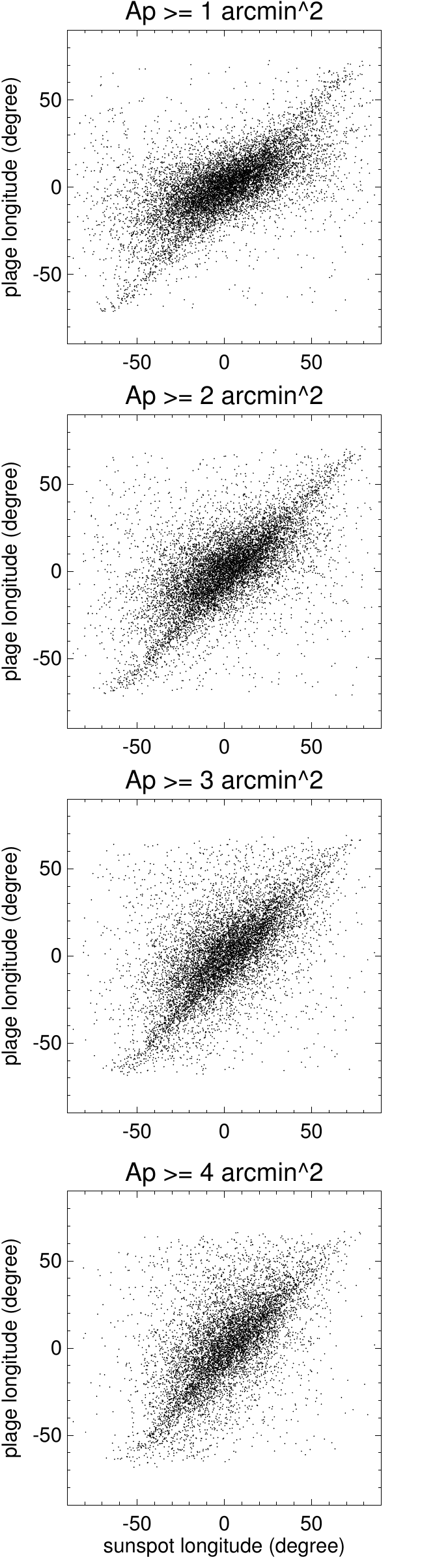}
     }
\caption{[Top to bottom]: Comparison of the `butterfly diagram' for the plages (black circles) and the sunspots (green circles). The longitudinal scatter diagram between the two are shown in the side panels. Different plage sizes (A$_{p}$) are mentioned in the title of the panels. An animated GIF of the above figure is available at \url{ftp://ftp.iiap.res.in/dipu/plage\_sunspot.gif}.}
\label{butter_plot}
\end{figure*}

Plages may be or may not be always associated with a sunspot due to the reasons mentioned in section 3.1. To probe this into depth, we divide the plages into four classes according to their individual sizes (A$_{p}$) and make use of the `Butterfly diagram' for our further analysis. In different panels in Figure~\ref{butter_plot} we plot the time-latitude diagram for the individually detected plages with black circles whereas the green circles represent the locations of the sunspots. In the first size class (where A$_{p}\geq$1 arcmin$^{2}$), we notice that there are substantially large number of plages which do not have any sunspots associated with them. Also we notice that at the end of every cycle (or may be from the next cycle due to the overlapping period) large number of plages appear at high latitudes ($\approx$60$^\circ$) and this is much more prominent in the southern hemisphere.\\ 
Now as we progressively go towards higher size thresholds (A$_{p}\geq$2 arcmin$^{2}$, A$_{p}\geq$3 arcmin$^{2}$, A$_{p}\geq$4 arcmin$^{2}$) we see that the two butterfly diagrams (one for the sunspots and the other for the plages) show progressively better match with each other. Thus, we conclude that the latitudinal locations of the plages with an area $\geq$4 arcmin$^{2}$ is showing a very good match with the sunspot latitudes.

A latitudinal match does not necessarily imply an one-to-one correspondence between the two. This is because these features can be at same latitudes but in different longitudes implying the two of them not connected at all. To better establish the association of plage sizes with the sunspot locations, we plot the longitudinal scatter diagrams for every plage size range as shown in the side panels in Figure~\ref{butter_plot}. We record the area weighted average longitudes for both the plages and the sunspots respectively during the simultaneous observing days. A careful analysis reveals that the scatter plots between the plage and the sunspot longitudes become progressively linear as we consider higher plage areas. Thus, in combination with our previous results from the `butterfly diagram' analysis, we conclude that plages with area $\geq$4arcmin$^{2}$ have better one to one correspondence with the sunspot locations. We must again remind the reader that the decay of a sunspot does not necessarily mean the disappearance of the associated plage.\\

 \subsection{Plage Sizes and The 11-Year Cycle}

\begin{figure}[!htbp]
\centering
\includegraphics[width=0.49\textwidth]{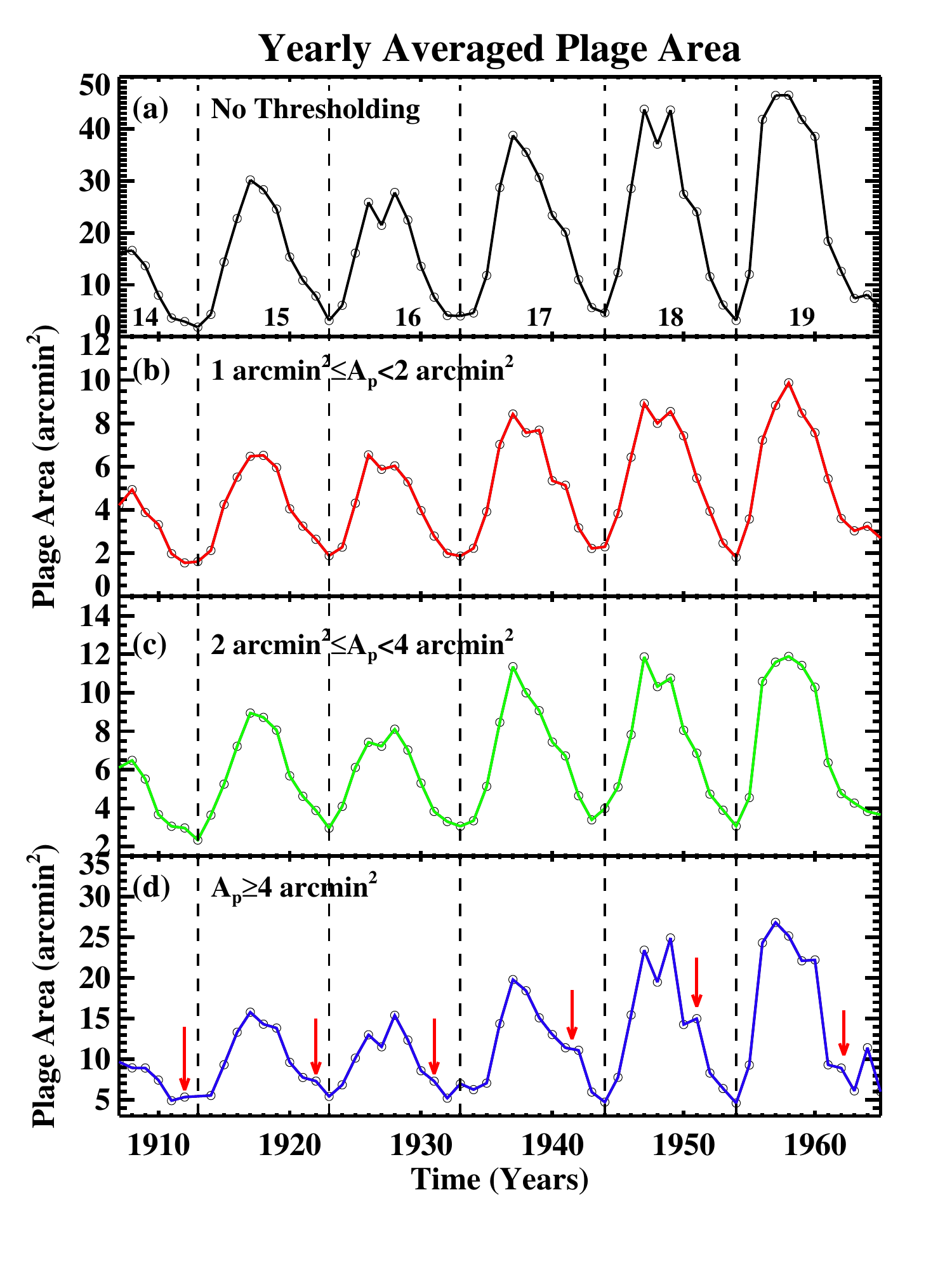}
\caption{Different panels showing the yearly averaged plage area time series as obtained using different plage size criteria. Individual size ranges are printed
on every panel.}
\label{size_plot} 
\end{figure}

Different long-term and short-term properties of the sunspot cycle show a strong size dependence \citep{2016ApJ...830L..33M}. Inspired by the findings of the plage size dependence on the plage-sunspot association in the previous section, we look for different signatures in the cycle properties when the plages are considered according to their sizes. In panel (a) of Figure~\ref{size_plot} we plot the `no thresholding' case whereas in panels (b-d) we plot the yearly averaged plage area variations for different size thresholds.

For the smallest sized plages (1arcmin$^{2}\leq$A$_{p}$$<$2 arcmin$^{2}$) we do not see much differences from the overall cycle characteristics (as found in panel \ref{size_plot}a), except for the double peak behavior for certain cycles. For an example, the double peaks of cycle 16 and cycle 18 become less prominent or weaker. Also the overall strength of the cycle do not change much for the smallest plage size range i.e the 19$^{th}$ cycle is still the strongest during the analyzed time span. The scenario almost remains the same for the medium plage size range (panel c). As we move towards the biggest plage sizes (panel d, we notice that the double peaks near the cycle maxima appear almost for every cycle. This is consistent with the behavior of biggest sized sunspots found by \citet{2016ApJ...830L..33M}. Apart from that we also note an interesting behavior i.e the presence of a weaker peak near the cycle minima. This is highlighted in the panel by red arrows. The location of this peak from the cycle maxima seems to move inwards as we move from cycle 14 to cycle 19. This trend is though not prominently visible for the 19$^{th}$ cycle as we see a double peak late after the cycle maxima. 

Next, we investigate the `odd-even rule' or the `G-O rule' which states that the odd numbered cycles are stronger than the preceding even numbered one \citep{1948..Astron..Zh..25..18G}. This is well established for the sunspot area and the sunspot number data. From Figure~\ref{size_plot} we notice that this rule is also valid for plage area time series for all the size ranges. Still it should me mentioned here that the relative heights of the cycles change slightly when one considers different size ranges (panels (b-d) of Figure~\ref{size_plot}).
\section{Summary And Conclusion}
Long term multi wavelength study of different solar features help us to better understand the magnetic field evolution in different layers of the solar atmosphere in different time scales. Sunspot and plages, though formed at different heights in the solar atmospheres, show a good correlation with each other at the time scales comparable to the solar cycle. In shorter time scales of months or days, we  find certain differences though. Such differences can be explained by considering the complex evolution of the magnetic fields associated with sunspots. When a sunspot decay the fields get fragmented and this progress results in disappearance of the same from the white light images. The left over small scale fields still live for quite a few days and continue to appear as plages in the subsequent Ca $\scriptsize{{\textrm{II}}}$ K images.

Analyzing the individual plages area we find that they follow a lognormal distribution similar to the sunspots. We also obtain a Gaussian distribution, in each of the hemispheres, of their latitudinal appearances. Some of the properties of the fitted Gaussian parameters show different evolution with the solar cycle as compared to the sunspots. Though not well understood but this hints towards a small scale component of the solar dynamo which is responsible for the evolution of the small scale fields. This aspect is explored further by implementing different size criteria on the individual plages and considering their time evolution. We find that different properties of the cycle change with the plage sizes which again points towards a complex dynamo operating into the Sun. Finally we use the `butterfly diagram' along with longitudinal scatter plots to show that the plages with sizes $\geq$ 4 arcmin$^{2}$ are always associated with a sunspot.\\

To conclude, we have analyzed the newly digitized Ca $\scriptsize{{\textrm{II}}}$ K data from Kodaikanal observatory to investigate the association of the plages, a chromospheric structure, with the sunspots which are the photospheric structures. To our knowledge, for the first time, individual plage sizes are considered and compared with the sunspot area data. Our analysis shows that the two layers (chromosphere and the photosphere) are magnetically coupled and the dynamo responsible for the magnetic fields in the Sun may have a complex action (generation of large scale and small scale fields).

\section{Acknowledgment}
{ We would like to thank the Kodaikanal facility of Indian Institute of Astrophysics, Bangalore, India for providing the data. This data is now available for public use at \url{http://kso.iiap. res.in/data}.}\\

 \bibliographystyle{apj}

\end{document}